# Electrically Reconfigurable Optical Metamaterial Based on

# Colloidal Dispersion of Metal Nano-Rods in Dielectric Fluid

Andrii B. Golovin[1] and Oleg D. Lavrentovich[1,2]

[1]Liquid Crystal Institute and [2]Chemical Physics Interdisciplinary Program,

Kent State University, Kent, Ohio, 44242.

E-mail: olavrent@kent.edu

*Optical metamaterials capture the imagination with breathtaking promises of nanoscale resolution in imaging and invisibility cloaking. We demonstrate an approach to construct a metamaterial in which metallic nanorods, of dimension much smaller than the wavelength of light, are suspended in a fluid and placed in a nonuniform electric field. The field controls the spatial distribution and orientation of nanorods because of the dielectrophoretic effect. The field-controlled placement of nanorods causes optical effects such as varying refractive index, optical anisotropy (birefringence), and reduced visibility of an object enclosed by the metamaterial.*

Optical metamaterials represent artificial composites with metal and dielectric elements intertwined at subwavelength scale. Different spatial architectures lead to fascinating optical properties such as negative refraction, subwavelength imaging, and cloaking [1-5]. Typically, the metastructures are fabricated by a nano lithography approach that has limited applicability when complex three dimensional (3D) arrangements or switching are required. So far, only two-dimensionsl (2D) non-switchable structures have been produced [4,5].

We demonstrate that 2D and 3D reconfigurable optical metamaterials can be produced by applying a non-uniform AC electric field to a dispersion of gold (Au) nanorods (NRs) in an isotropic dielectric fluid. Previously, manipulation in a non-uniform field, also known as dielectrophoretic effect [6], was demonstrated for metal nanowires of much larger supra-micron length, see, e.g., [7-11]. We show that the dielectrophoretic control of NRs that are only (50-70) nm long, leads to optical effects such as spatially varying refractive index, birefringence and reduced visibility of an object enclosed by the metamaterial.

We used two sets of Au NRs: (1) "long" NRs with average diameter $d = 12$ nm, length $L = 70$ nm, and an absorption peak due to the longitudinal plasmon at ~1050 nm and (2) "short" NRs, $d = 20$ nm, $L = 50$ nm, and an absorption maximum at ~725 nm. The short NRs are well suitable to explore the spatial distribution and orientation of NRs through light absorption while the long NRs are better suited to observe cloaking and birefringence effects in the visible part of spectrum. The NRs, functionalized with polystyrene (PS) [12, 13] were dispersed in toluene with a refractive index $n_t = 1.497$ at $\lambda = 589\,\text{nm}$. The volume fraction of NRs was (4-8) $\times 10^{-4}$. We experimented with flat and cylindrical cells; the latter produce 3D configurations similar to the optical cloak of Cai et al. [3].

(**1**). **A flat cell** is formed by two glass plates, confining two mutually perpendicular electrodes, Figs.1. The ground electrode (3) is a copper wire of diameter $2a{=}3\,\mu\mathrm{m}$ in a borosilicate glass shell of diameter $20\,\mu\mathrm{m}$ that determines the separation between the glass plates. The second electrode (2) is a similar wire, connected to the waveform generator, with the glass shell etched out along the last portion, about 1 mm long, near the electrode's tip. The cell is filled with the dispersion of Au NRs in toluene and sealed.

At zero voltage, the NRs are distributed uniformly across the area and show no alignment. In *crossed* polarizers, the texture is dark, Fig.1(b). When the AC voltage $U{=}170\,\mathrm{V_{rms}}$, frequency $f=100$ kHz is applied, the Au NRs, being more polarizable than toluene, move into the regions of high electric field and align, creating an optically birefringent cloud near the electrode (2), Fig.1(d). By inserting an optical compensator one establishes [14] that in the birefringent region, the index of refraction for light polarized perpendicular to the electrode (2) is smaller than for parallel polarization, $\Delta n = n_{\parallel} - n_{\perp} < 0$, consistent with the alignment of NRs along the field, Fig.1(c). The latter is also confirmed by absorption of linearly polarized light near the peak of longitudinal plasmon absorption.

To characterize the concentration variation of NRs along the axis $x$ crossing the electrode (2) near its tip, Fig.1(d), we measured the transmittance $T_{\parallel}$ of linearly polarized (along $x$) light as a function of $x$. The wavelength $\lambda$ was chosen at 460 nm, for which the anisotropy of light absorption is small, so that the variation of $T_{\parallel}$ is determined mostly by the concentration gradients. We determine the ratio $k(x) = \dfrac{\eta_U(x)}{\eta_0} = \dfrac{\ln T_{\parallel}(x)}{\ln T_{\parallel,0}}$ as the measure of how much the local filling factor $\eta_U(x)$ of Au NRs in the field is larger than the filling factor $\eta_0 = const$ in zero field. The spatial variation $k(x)$ clearly indicates that the NRs are accumulated in the region of high field, Fig.2(a). The maximum $k$ is ~20, corresponding to $\eta_U \sim 0.01 - 0.02$, Fig.2(a) [15].

To quantify birefringence near the electrode, $\Delta n_{\max}$, we selected a small circular region of diameter 5 μm (marked in Fig.1(d)), centered at the point of the maximum $k(x)$ and measured transmission of light polarized parallel to the $x$ axis ($T_{||}$), perpendicular to it ($T_{\perp}$) and at the angle of 45 degrees ($T_{45}$), as the function of $\lambda$, and calculated the quantify

$$\Phi = \frac{\lambda}{2\pi} \cdot \cos^{-1} \frac{4T_{45} - (T_{II} + T_{\perp})}{2\sqrt{T_{II} T_{\perp}}}$$, Fig.2(b). $\Phi$ is a quantitative equivalent of the true optical

retardation of an absorbing material [16] ($\Phi$ would represent a true retardation if all NRs are aligned in the plane of the cell). $\Phi$ is significant, reaching (-190) nm at 650 nm, Fig.2(b). By approximating $\Phi$ as $h\Delta n_{\max}$, where $h \approx$ (2-4) μm is the effective pathway of light trough the assembled NRs, one estimates $\Delta n_{\max} \sim \Phi / h \sim -(0.1 \div 0.05)$ at $\lambda = 650$ nm , comparable to birefringence of liquid crystals [14]. When the electric field is on, $\Delta n(x)$ changes from $\Delta n = 0$ at large distance $\delta x \geq 10$ μm from the electrode surface, to $\Delta n_{\max} \sim -(0.1 \div 0.05)$ at $\delta x \approx 1$ μm; for smaller $\delta x < 1$ μm, the value of $\Delta n(x)$ might drop again because of the apparent depletion effect [15]. In the field off state, $\Delta n = 0$ for all distances $\delta x$. The gradients created by the field-induced concentration and orientation of NRs near the central electrode are sufficient to cause experimentally observed effects in the cylindrical cells described below.

(**2). Cylindrical capillary.** A cavity of diameter 14 μm in a cylindrical glass capillary of diameter 57 μm is filled with a dispersion of Au NRs, Figs.3 and 4. The electric field is created by coaxial cylindrical electrodes, one being a copper wire (2) of diameter $2a = 3\,\mu m$ running along the axis and the second one formed by a transparent indium tin oxide (ITO) deposited on the external surface (3) of the capillary. The radial field $E_r \propto 1/r$ decreases with the distance $r > a$ from the wire. The field accumulates and aligns the NRs near the central electrode (2), Figs.3 (b) and 3 (c).

The most striking optical feature is that the applied field weakens the shadow of non-transparent central electrode (2), observed in the orthoscopic mode under the microscope with parallel polarizers. The effect is wavelength and polarization dependent. Fig.3(d) shows light transmittance for the red component of the RGB signal measured by the video camera, $\lambda = (550\text{-}700)\,nm$. The shadow reduction is noticeable for light polarized perpendicularly to the capillary, see Fig. 3(d) and video Fig.4 but not for parallel polarization, Fig.3(e). The effect in Fig.4 is an "imperfect" experimental version of the cloak model proposed by Cai et al. [3]. The local refractive index changes from a smaller value $n_\parallel(x) < n_\perp$ near the central electrode, to a larger value $n_\parallel \approx n_\perp \approx n_t$ at the periphery. This index gradient bends the light rays around the electrode, thus reducing its visibility. Propagation of light with parallel polarization is hardly affected by the electric field, as $n_\perp \approx n_t \approx const$.

The contribution of Au NRs to the effective $n_\parallel$ can be roughly estimated as $n_\parallel \approx \sqrt{\left(1 - \eta_U\right) n_t^2 + \eta_U\, n_{NR}^2}$, where $n_{NR}^2$ is the (real part of) dielectric permittivity of Au at optical frequencies, e.g., $n_{NR}^2 \approx -20$ at 700 nm [17] (NRs do not change $n_\perp \approx n_t$, since their response to the perpendicular light polarization is weak [3]). For $\eta_U \approx 0.01-0.02$, one finds $n_\parallel \approx 1.34-1.42$ and thus $\Delta n \approx -\left(0.16 \div 0.09\right)$, similar to the experimental data. If PS is aligned around the NRs, it can influence $\Delta n$, too. For stretched PS, $\left|\Delta n_{PS}\right| \approx 0.0006$ [18]. If the entire 20 μm thick cell is filled with such a birefringent PS, its phase retardation would contribute only about 12 nm to the much larger values of $\Phi$ in Fig.2(b).

Let us estimate the dielectrophoretic force acting on a NR. In dipole approximation [6],

$$F_{DEP} \approx \frac{\pi}{8} d^2 L \varepsilon_t \, \mathrm{Re}\left\{ \frac{\varepsilon_{NR}^* - \varepsilon_t^*}{\varepsilon_t^*} \right\} \nabla \left| E_{e,rms} \right|^2, \quad \text{where} \quad \pi d^2 L / 8 \quad \sim 10^{-23}\ \mathrm{m}^3, \quad \varepsilon_t = 2.4 \varepsilon_0,$$

$\varepsilon_0 = 8.854 \times 10^{-12}$ C/(Vm), $\mathrm{Re}\left\{ \dfrac{\varepsilon_{NR}^* - \varepsilon_t^*}{\varepsilon_t^*} \right\} = \dfrac{\omega^2 \left(\varepsilon_{NR} \varepsilon_t - \varepsilon_t^2\right) + \left(\sigma_{NR} \sigma_t - \sigma_t^2\right)}{\omega^2 \varepsilon_t^2 + \sigma_t^2}$ is the real part of the function of complex permittivities $\varepsilon^* = \varepsilon - i \sigma / \omega$ of the NR and the medium (subscripts "NR" and "t", respectively), $\sigma$ is conductivity, $\omega = 2\pi f$. With $\varepsilon_{NR} = 6.9 \varepsilon_0$, $\sigma_{NR} = 4.5 \times 10^7$ S/m [16], $\sigma_t \sim 5 \times 10^{-11}$ S/m [19], $f = 10^5$ Hz, $E_e \sim 10^6$ V/m, and the scale of gradients $\left(10-100\right)$ μm, one estimates $F_{DEP} \sim \left(10\text{-}100\right)$ pN, above the random forces of Brownian nature, $F \sim \dfrac{k_B T}{d} \sim \left(0.1\text{-}1\right)$ pN at room temperature $T$. Note that field-induced condensation of NRs takes place over extended regions of space; it is also reversible, as the NRs randomize their orientation and position once the field is switched off. The structures in Figs.1(d) and 3(c)-3(e) are stationary: once established within a few seconds after the field is applied, they do not evolve if the field is kept constant. These features suggest that the NRs repel each other. The natural mechanisms are steric and electrostatic; they exist even when the field is zero. An interesting source of repulsion is the external field itself: the field-induced dipoles in NRs repel each other if the NRs are located side-by-side.

Experiments above demonstrate that a non-uniform electric field applied to a colloidal dispersion of submicron Au NRs, is capable of concentrating the particles in the region of the maximum field and also to align them parallel to the field. The effect induces gradient refractive index for polarized light that is decreasing from the high-field region to the low-field region. In the cylindrical sample, the effect represents an imperfect experimental realization of the theoretical cloak model proposed by Cai et al. [3], as evidenced by a mitigated shadow of a non-transparent object, in our case the central electrode (2), see Figs.3, 4 and Ref. [20].

This work was supported by DOE DE-FG02-06ER46331 and AFOSR MURI FA9550-06-1-0337 grants. We thank N.A. Kotov and P. Palffy-Muhoray for providing us with Au NRs dispersions; A. Agarwal, J. Fontana, P. Luchette, H.-S. Park, B. Senyuk, H. Wonderly, and L. Qiu for help in sample preparations. We thank P. Palffy-Muhoray, V. M. Shalaev, C. Y. Lee, A. V. Kildishev, S.V . Shiyanovskii and V. P. Drachev for fruitful discussions.

**References**

1.  J. B. Pendry, D. Schurig, D. R. Smith, Science **312,** 1780 (2006).

2.  U. Leonhardt, Science **312,** 1777 (2006).

3.  W. Cai, U. K. Chettiar, A. V. Kildishev, V. M. Shalaev, Nature Photonics **1,** 224 (2007).

4.  I. I. Smolyaninov, Y. J. Huang, C. C.  Davis, Optics Letters **33,** 1342 (2008).

5.  J. Valentine, J.S. Li, T. Zentgraf, G. Bartal, X. Zhang, Nature Materials **8**, 568 (2009).

6.  H. Morgan, N. G. Green, *AC Electrokinetics: Colloids and nanoparticles* (Research Studies Press Ltd., Baldock, England (2003).

7.  P. A. Smith, C. D. Nordquist, T. N. Jackson, T. S. Mayer, B. R. Martin, J. Mbindyo, and T. E. Mallouk, Appl. Phys. Lett. **77,** 1399 (2000)

8.  H. W. Seo, C. S. Han, S. O. Hwang, J. Park, Nanotechnology **17,** 3388 (2006).

9.  S. J. Papadakis, Z. Gu, D.H. Gracias, Appl. Phys. Lett. **88,** 233118 (2006).

10.  B. Edwards, N. Engheta, S. Evoy, *J.* Appl. Physics **102,** 024913 (2007).

11.  J.J. Boote, S.D. Evans, Nanotechnology **16,** 1500 (2005).

12.  J. Fontana, A. Agarwal, N. Kotov, and P. Palffy-Muhoray, American Physical Society March Meeting, Pittsburgh, PE, 16 March 2009.

13. Z. Nie, D. Fava, E. Kumacheva, S. Zou, G.C. Walker, M. Rubinstein, Nature Materials **6,** 609 (2007).

14. M. Kleman, O. D. Lavrentovich, *Soft Matter Physics: An Introduction* (Springer New York, 2003), pp. 96-98.

15. Interestingly, $k(x)$ decreases near the very surface of electrodes, which might indicate a NR-depleted thin layer associated with osmotic or electrostatic surface effects.

16. Yu. A. Nastishin, H. Liu, T. Schneider, V. Nazarenko, R. Vasyuta, S. V. Shiyanovskii, O. D. Lavrentovich,. Phys. Rev. E **72,** 041711 (2005).

17. I. El-Kady, M. M. Sigalas, R. Biswas, K. M. Ho, C. M. Soukoulis, Phys. Rev. B **62,** 15299 (2000).

18. M. Jiao, S. Gauza, Y. Li, J. Yan, S. T. Wu, T. Chiba, App. Phys. Lett. **94,** 101107 (2009).

19. M. V. Sapozhnikov, Y. Tolmachev, I. S. Aranson, W.K. Kwok, Phys. Rev. Lett. **90,** 114301 (2003).

20. See EPAPS supplementary material at *A. B. Golovin, O. D. Lavrentovich, Appl. Phys. Lett. 95, 254104 (2009)* for the detailed experimental procedures and electrically controlled optical effects.

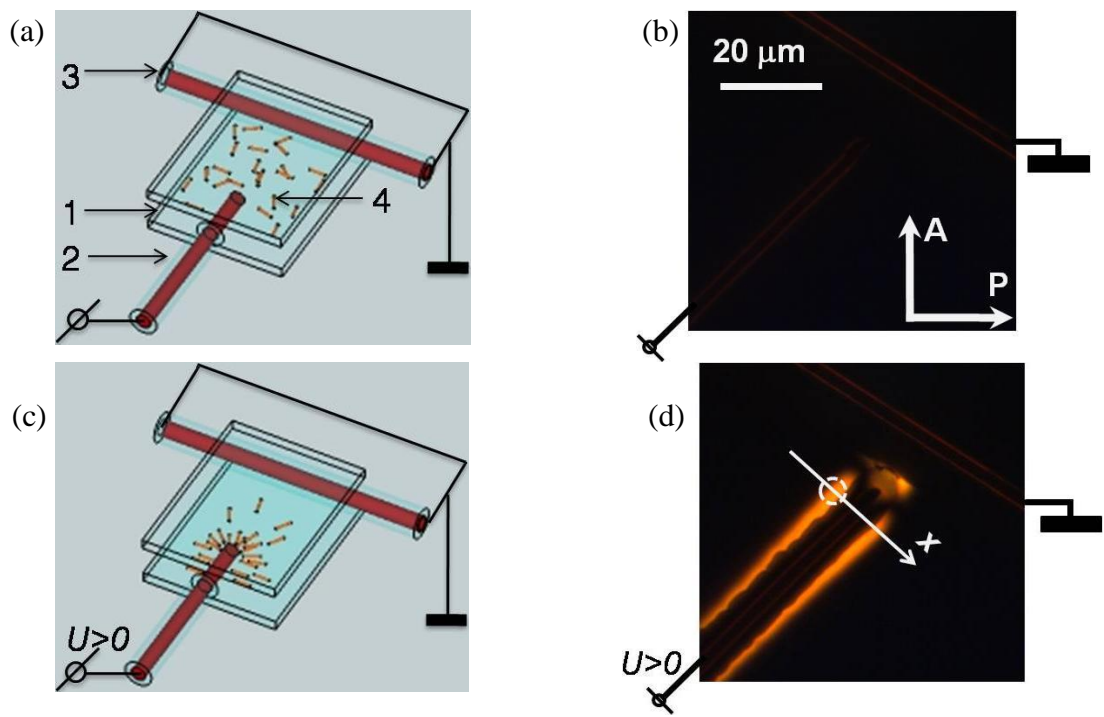

Fig.1. (Color online) Flat glass sample (1) with two electrodes (2,3), filled with "short" Au NRs dispersed in toluene (4). At zero field, the dispersion is isotropic (a) and produces a dark texture (b) under the microscope with crossed polarizers A and P. When the voltage is on ($U = 170 \, \mathrm{V_{rms}}$, $f = 100 \, \mathrm{kHz}$), a birefringent cloud of aligned NRs accumulates near the electrode (2) (c,d).

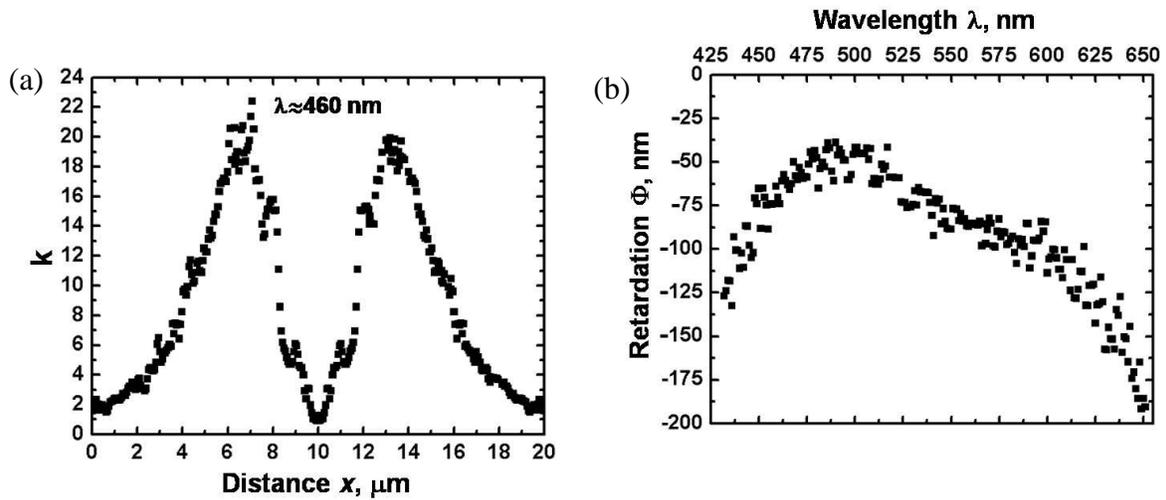

Fig.2. (a) Applied voltage $U = 170\,\mathrm{V_{rms}}$, $f = 100\,\mathrm{kHz}$ causes an increase in the filling factor $\boldsymbol{\kappa}$ that changes along the axis x marked in Fig.1(d); data are presented for "short" NRs; (b) the "long" NRs assembly is birefringent near the electrode (2); the effective optical phase retardation $\Phi$ is measured as a function of $\lambda$ in the circular region marked in Fig.1(d).

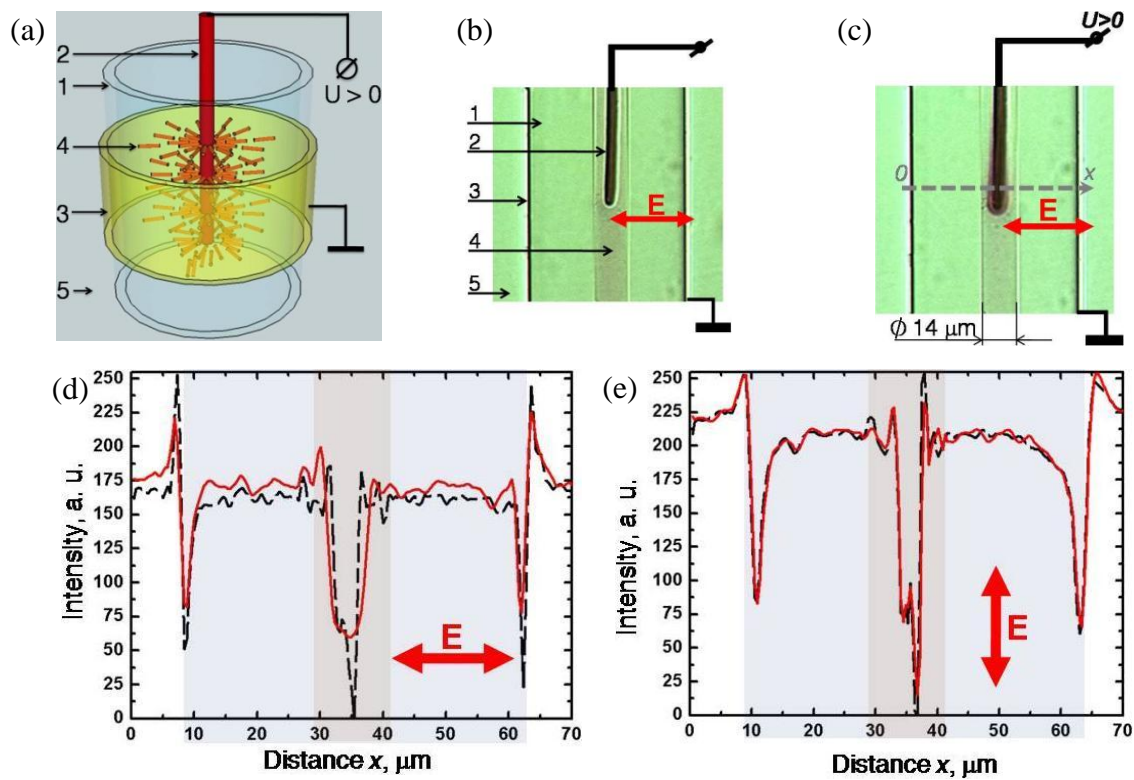

Fig.3. (Color online) (a) Cylindrical glass capillary (1) with an axial copper wire electrode (2) and a transparent electrode at the outer surface (3), filled with "long" Au NRs in toluene (4), and fixed in polymerized optical adhesive (5). Microscope textures (parallel polarizers) of a capillary filled with the long NRs when the field is off (b) and on, $U = 170\,\mathrm{V_{rms}}$, $f = 100\,\mathrm{kHz}$ (c). Electric field-induced redistribution of Au NRs changes the profile of light transmission through the capillary for the light polarization **E** perpendicular to the capillary (d), but not for **E** parallel to the capillary (e). Red trace: field on, black dashed trace: field off.

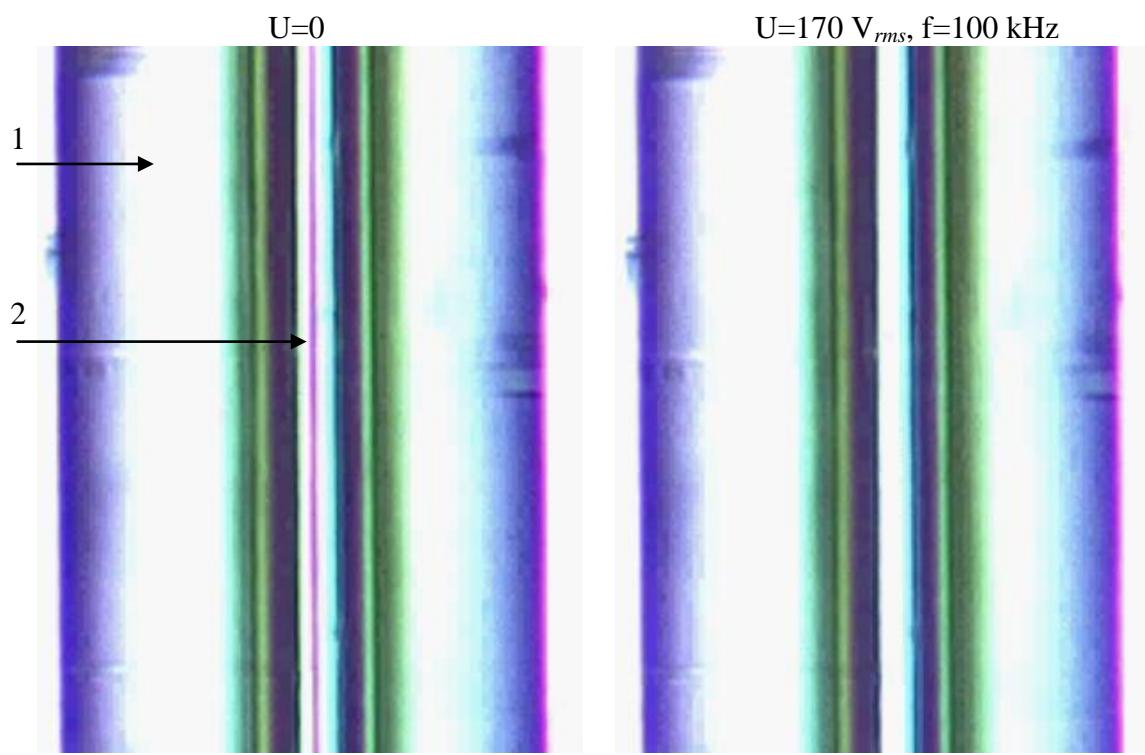

Fig.4. (Color online) Two frames of a video file demonstrating periodic change in visibility of the central electrode (2) in a cylindrical capillary (1) filled with dispersion of "short" Au NRs in toluene, under the applied voltage 170 $V_{rms}$, 100 kHz, modulated with a frequency of 0.5 Hz. Observation under the microscope with light polarized normally to the capillary axis. The corresponding video file is available at aip.org.

**Supplementary Information**

**Materials and Methods**

We used borosilicate glass round capillaries (GW Lab, Canoga Park, California), refractive index $n = 1.517$ at $\lambda = 588$ nm (close to $n_t$), with outer diameter 57 $\mu$ m and inner diameter 14 $\mu$ m. A thin conductive layer of ITO was deposited at the outer surface (Genvac AeroSpace Inc.). The second electrode was a copper wire of diameter 3 $\mu m$ (GW Lab) inserted into the capillary along the axis. The sample was fixed at 1 mm thick borosilicate glass substrates by a clear optical adhesive NOA76 (Norland) with a matching refractive index $n = 1.51$. The ends of capillary were sealed by epoxy glue to slow down evaporation of toluene.

Nikon polarizing microscope Eclipse E 600, Nikon objective (Plan Apo 60xA/1.40 Oil, DIC H, /0.17, WD 0.21), CCD camera Hitachi HV-C20U-S4, and Nikon photometry system G70 with photo-detector P100S were used for optical measurements. Images captured by CCD camera were analyzed with a computer program Image-Pro Plus 6.2 (Media Cybernetics Inc.). Arbitrary waveform generator Keithley 3390 50 MHz and wideband amplifier Krohn-Hite 7602M were used to apply AC sinusoidal voltage waveform.

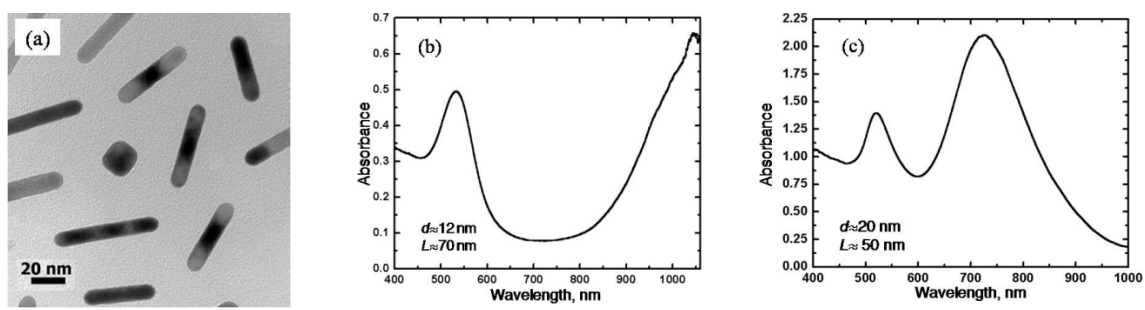

Fig.S1. Transmission electron microscope image of dried "long" Au NRs (a). Absorption spectra of toluene dispersions of "long" (b) and "short" (c) Au NRs. The spectrum (b) was measured by Dr. P. Luchette.

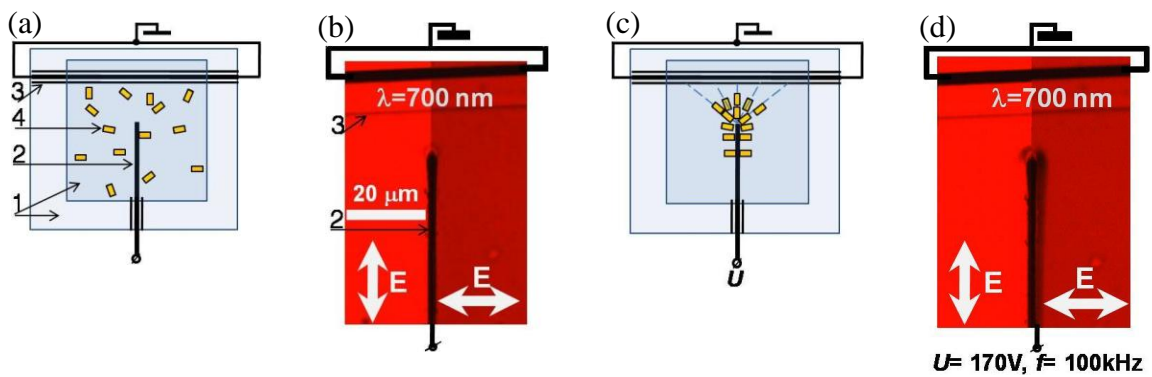

Fig.S2. (Color online) A flat glass cell (1) with two mutually perpendicular electrodes (2, 3), filled with "short" Au NRs dispersed in toluene (4) when the field is off (a, b) and on (c, d). Orientation of Au NRs is random when the field is off (a, b). When the field is on ( $U = 170$ V$_{rms}$, $f = 100$ kHz ), the NRs accumulate in the regions of the highest field, near the vertical electrode (c, d). The effect is evident when the cell is observed in linearly polarized light with wavelength 700 nm, close to the longitudinal absorption peak for "short" NRs, see Fig. S1(c). When the field is on, and the polarization **E** of light is horizontal, the condensed and aligned NRs are seen as a dark cloud around the vertical electrode, right part of (d). Only a portion of this cloud is visible when **E** is vertical, left part of (d); since **E** is normal to most of NRs in this case. No clouds of NRs are observed when the field is off, regardless of light polarization (b).

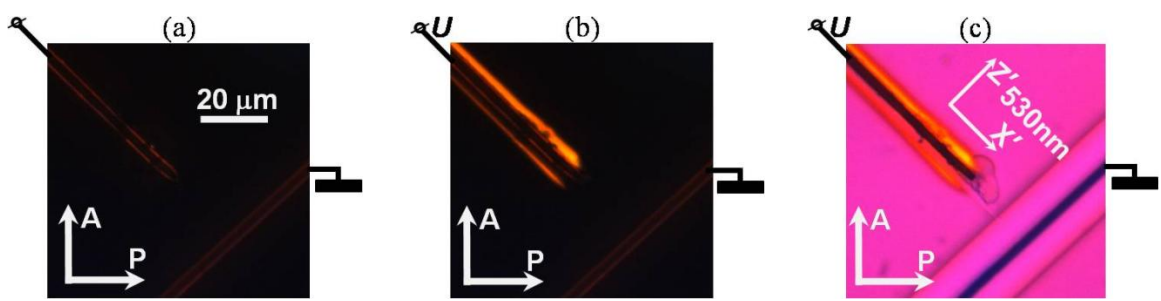

Fig.S3. (Color online) (a) Polarizing microscopy texture of a flat cell when there is no field; the "short" Au NRs dispersion is isotropic and appears dark between the crossed analyzer and polarizer. The applied electric field creates a birefringent zone around the electrode (b,c), as observed in the regime of crossed polarizers (b) or crossed polarizers with an optical compensator, a 530 nm waveplate (c).

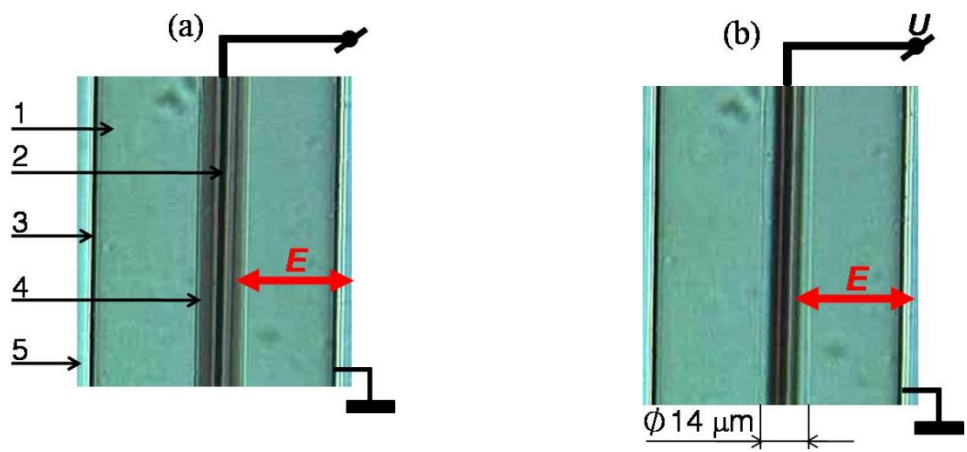

Fig.S4. (Color online) Polarizing microscopy of a capillary similar to the one in Fig.3 but filled with "short" Au NRs when the electric field is off (a) and on (b).

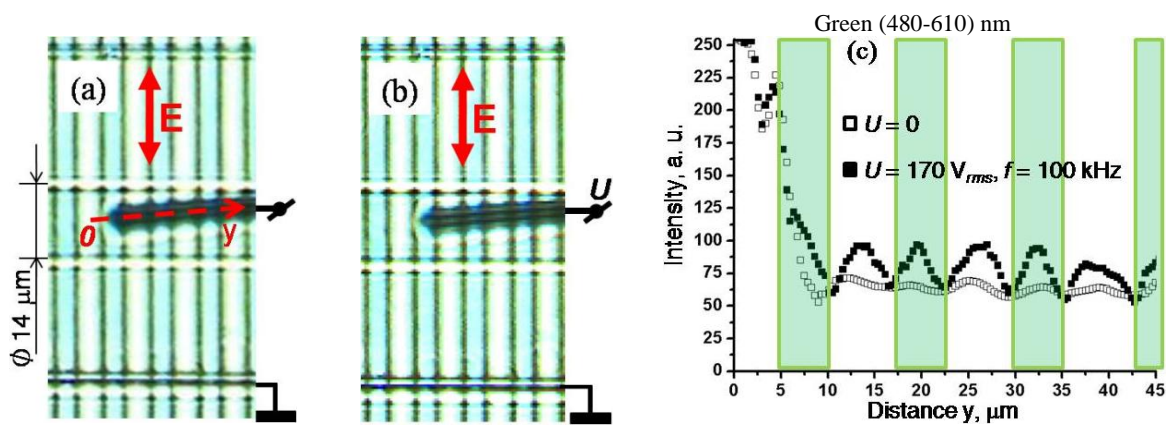

Fig.S5. (Color online) The effect of *enhanced* visibility of an object placed behind the switchable metamaterial. The object is a stripe pattern of a cured photoresist Shipley S1818 on a flat glass plate. The capillary with toluene dispersion of "long" Au NRs is placed directly on top of it. The axis $y$ is drawn along the axis of the electrode. The stripe pattern is observed through the capillary, with linearly polarized light, under an optical microscope, when the voltage is off (a) and on, $U = 170$ V$_{rms}$, $f = 100$ kHz (b). At zero voltage, the central electrode (2) blocks the image of the stripes beneath it (a). However, once the field is applied, the visibility of stripes is enhanced (b,c), as evidenced by the profile of light transmittance along the axis of electrode (c). Similarly to the experiment in Fig.3, the gradient refractive index bends the optical rays around the obstacle (the central electrode), reducing its shadow.